\documentclass[aps,prc,twocolumn,showpacs,preprintnumbers,nofootinbib,float,superscriptaddress,longbibliography]{revtex4-2}
\usepackage{graphicx}
\usepackage{color}
\usepackage[dvipsnames]{xcolor}
\usepackage[colorlinks=true, pdfstartview=FitV, linkcolor=RedOrange, citecolor=ForestGreen, urlcolor=blue]{hyperref}
\usepackage{amsmath}
\usepackage{amssymb}
\usepackage{bm}
\usepackage{physics}
\usepackage{placeins}
\usepackage{adjustbox}
\usepackage{comment}

\newcommand{\btheta}{{\boldsymbol{\theta}}}
\newcommand{\bomega}{{\boldsymbol{\omega}}}
\newcommand{\by}{\mathbf{y}}

\begin{document}

\title{Learning Informed Prior Distributions with Normalizing Flows for Bayesian Analysis}

\author{Hendrik Roch} \email{Hendrik.Roch@wayne.edu}
\affiliation{Department of Physics and Astronomy, Wayne State University, Detroit, Michigan 48201, USA}
\author{Chun Shen} \email{chunshen@wayne.edu}
\affiliation{Department of Physics and Astronomy, Wayne State University, Detroit, Michigan 48201, USA}

\begin{abstract}
We investigate the use of normalizing flow (NF) models as flexible priors in Bayesian inference via Markov Chain Monte Carlo (MCMC) sampling for iterative Bayesian calibration. 
Trained on posteriors from previous analyses, these models can be used as informative priors that capture non-trivial distributions and correlations in subsequent inference tasks. 
We compare different training strategies and loss functions, finding that training based on Kullback–Leibler (KL) divergence and unsupervised learning consistently yield the most accurate reproductions of reference distributions.
We apply such a sequential Bayesian workflow to a high-energy nuclear physics problem; MCMC with NF-based priors reproduces the results of one-shot joint inference well, provided the target distributions are unimodal. 
In cases with pronounced multi-modality or dataset tension, distortions may arise, underscoring the need for caution in multi-stage Bayesian inference. 
A comparison between the \texttt{pocoMC} MCMC sampler and the standard \texttt{emcee} sampler further demonstrates the importance of advanced and robust algorithms for exploring the posterior space. 
Overall, our results establish NF-based priors as a practical and efficient tool for sequential Bayesian inference in high-dimensional parameter spaces.
\end{abstract}

\maketitle

\section{Introduction}
\label{sec:intro}
Bayesian inference~\cite{10.1093/oso/9780198568315.001.0001} is a systematic statistical framework to constrain the probability distributions of model parameters $\btheta$, based on comparisons between model predictions $\by(\btheta)$ and experimental data $\by_{\rm exp}$. 
This framework can naturally handle high-dimensional model parameter spaces and apply multiple experimental constraints with non-trivial covariances to theoretical models~\cite{RevModPhys.83.943, 10.1214/10-BA524, Phillips:2020dmw}
It has become a standard data-driven approach in high-energy nuclear physics~\cite{Paquet:2023rfd, Bernhard:2019bmu, Nijs:2020roc, Parkkila:2021tqq, Heffernan:2023gye, Shen:2023awv, Jahan:2024wpj, JETSCAPE:2024cqe}. 
The resulting multidimensional posterior distributions for the model parameters can be used to propagate uncertainties to model predictions~\cite{Nijs:2021clz, Jahan:2025cbp, Wu:2025psu} in nuclear physics, under the scope of Bayesian uncertainty quantification (BUQ)~\cite{Phillips:2020dmw, Jacobs:2025ncn, Jaiswal:2025hyp}.

At its core, Bayesian inference relies on Bayes' theorem:
\begin{align}
\mathcal{P}(\btheta|\mathbf{y}_{\rm exp}) = \frac{\mathcal{P}(\mathbf{y}_{\rm exp}|\btheta)\mathcal{P}(\btheta)}{\mathcal{P}(\mathbf{y}_{\rm exp})},
\label{eq:Bayes_theorem}
\end{align}
where $\mathcal{P}(\btheta|\mathbf{y}_{\rm exp})$ is the posterior distribution, $\mathcal{P}(\mathbf{y}_{\rm exp}|\btheta)$ the likelihood, $\mathcal{P}(\btheta)$ the prior, and $\mathcal{P}(\mathbf{y}_{\rm exp})$ the evidence.

A key ingredient is the prior distribution $\mathcal{P}(\btheta)$, which encodes prior knowledge about model parameters.
Typically, uniform priors are used to express unbiased prior preference, or uncorrelated Gaussian priors are applied when parameters are expected to cluster around known values.
These distributions are easy to implement and sample from during Markov Chain Monte Carlo (MCMC) analysis, making them popular in practice.

However, incorporating informative priors is crucial when conducting sequential Bayesian analysis to investigate how different sets of experimental data progressively influence the posterior distribution. For example, in high-energy nuclear physics, a ``multi-messager'' Bayesian analysis using experimental data from different fields, such as Deep Inelastic Scatterings (DIS) and heavy-ion collisions~\cite{Mantysaari:2025ltq, Mantysaari:2025tcg, Andronic:2025ylc}, would set strong constraints on the initial-state parameters and the properties of Quark-Gluon Plasma. Moreover, integrating knowledge from previous Bayesian analyses into the prior distribution can improve efficiency in new inference tasks by significantly reducing the volume of parameter phase space. This optimization is critical when the new observables are computationally expensive. By shrinking the prior parameter space, the sequential Bayesian inference enables us to efficiently emulate statistically demanding observables.

Yet, using posterior distributions from earlier analyses directly as priors can be challenging.
They may be multimodal, non-Gaussian, or concentrated away from the center of the uniform prior, and often encode non-trivial correlations between parameters that are difficult to represent analytically and to sample from efficiently with conventional methods.

One strategy would be to draw samples from an ensemble chain generated in the previous MCMC analysis as the prior for the subsequent study. It becomes impractical as the dimension of the model parameter increases, and one is limited to discrete sample points rather than a continuous distribution.
Alternatively -- and more flexibly -- one can train a generative model, such as a normalizing flow (NF), on these samples~\cite{Yamauchi:2023xrz}.
The trained NF model can then produce new samples efficiently, while preserving complex structures of the original distribution, including parameter correlations. This approach is particularly valuable in high-dimensional parameter spaces, where capturing correlations and non-standard shapes in the prior becomes essential for accurate inference.\footnote{High dimensionality in most high-energy nuclear physics applications is in the range of 20 to 50 parameters.}

Such an NF-based generative model has been developed in Ref.~\cite{Yamauchi:2023xrz} and tested on synthetic distributions of moderate dimension. In this work, we further extend this NF framework to unsupervised learning cases where posterior densities are unavailable. Moreover, we incorporate this NF-based model as an informative prior for a sequential Bayesian analysis in high-energy nuclear physics, where we apply constraints from different sets of experimental data in succession. We systematically verify the obtained posterior result with that by performing a one-shot joint Bayesian inference with all experimental constraints at once. We further explore the posterior consistency by switching the order in sequential Bayesian analysis.

The paper is organized as follows. In Sec.~\ref{sec:model}, we introduce the normalizing flow model and the Bayesian framework used to perform sequential Bayesian inference.
Section~\ref{sec:results} then applies this framework to a representative example in a seven-dimensional parameter space from a study in high-energy nuclear physics~\cite{Mantysaari:2025ltq}.
Finally, in Sec.~\ref{sec:conclusion}, we summarize our findings and outline future directions for applying and extending this approach.

\section{The Theoretical Framework}
\label{sec:model}
In this section, we present the theoretical and computational framework for multi-stage Bayesian inference, where the posterior distribution obtained from the first-stage Bayesian study is used as an informative prior for a subsequent analysis. This study adopts the NF model framework from Ref.~\cite{Yamauchi:2023xrz}. We extended it with an unsupervised learning capability to deal with distribution ensembles with unknown probability densities.

\subsection{Normalizing Flow Model}
\label{subsec:NFmodel}

An NF constructs a bijective mapping between a non-trivial target distribution $p(\btheta)$ in $\mathbb{R}^N$ and a simpler, usually Gaussian, reference distribution $p_G(\bomega)$ in $\mathbb{R}^N$~\cite{JMLR:v22:19-1028}.
Specifically, the NF $\mathcal{F}$ maps latent variables $\bomega$ to the original parameter space via $\btheta = \mathcal{F}(\bomega)$, such that:
\begin{align}
\mathrm{d}\btheta\; p(\btheta)
= \mathrm{d}\bomega\; \det\left(\frac{\partial\btheta}{\partial\bomega}\right) p_G(\bomega).
\label{eq:NFexact}
\end{align}
Here, $\det(\partial\btheta/\partial\bomega)$ is the Jacobian determinant of the transformation.

Such transformations exist for normalizable, non-negative distributions~\cite{Kobyzev:2019ydm}, though they may not be unique, especially in higher dimensions.

Once trained, the NF model enables efficient sampling from the target distribution $p(\btheta)$: samples of $\bomega$ are drawn from the multivariate Gaussian $p_G(\bomega)$ and mapped to $\btheta$ via $\btheta=\mathcal{F}(\bomega)$.
This also captures complex correlations between parameters that are typically difficult to encode explicitly in conventional priors.

In practice, the exact mapping in Eq.~\eqref{eq:NFexact} is infeasible for general high-dimensional distributions.
Therefore, an approximate mapping $\mathcal{F}:\bomega\rightarrow\btheta$ is learned, yielding an approximate distribution $p^\prime(\btheta)$:
\begin{align}
\mathrm{d}\bomega\; p_G(\bomega) = \mathrm{d}\btheta\; p^\prime(\btheta) \approx \mathrm{d}\btheta\; p(\btheta).
\end{align}

The similarity between $p(\btheta)$ and $p^\prime(\btheta)$ is quantified using Jeffreys' divergence:
\begin{align}
D_{\rm J}(p,p^\prime)
= \int\mathrm{d}\btheta\, \left[
\tilde{p}(\btheta)\ln\left(\frac{p(\btheta)}{p^\prime(\btheta)}\right)
+ \tilde{p}^\prime(\btheta)\ln\left(\frac{p^\prime(\btheta)}{p(\btheta)}\right)
\right],
\label{eq:loss_Jeffreys}
\end{align}
where the densities $\tilde{p}$ and $\tilde{p}^\prime$ are normalized distributions for $p(\btheta)$ and $p^\prime(\btheta)$, respectively. Alternatively, one can use the Kullback–Leibler (KL) divergence:
\begin{align}
    D_{\rm KL}(p,p^\prime)
= \int\mathrm{d}\btheta\,
\tilde{p}(\btheta)\ln\left(\frac{p(\btheta)}{p^\prime(\btheta)}\right).
\label{eq:loss_KL}
\end{align}

To further refine the approximation, a reweighting technique can be applied when sampling from the NF.
During training, this technique helps evaluate the loss function and improve agreement between $p(\btheta)$ and $p^\prime(\btheta)$ (see Ref.~\cite{Yamauchi:2023xrz} for details).

The NF is realized as a neural network based on the Real NVP (Real-valued Non-Volume Preserving) model~\cite{DBLP:DinhSB16}, followed by an additional scale-and-shift layer.
Schematically, the entire NF transformation $\mathcal{F}(\bomega)$ can be written as:
\begin{align}
    \mathcal{F}(\bomega) = L \circ \bigl( A_o \circ A_e \bigr)^\mathcal{N} (\bomega),
\label{eq:NF_architecture}
\end{align}
where $A_e$ and $A_o$ are affine coupling layers with even and odd masking, respectively, repeated $\mathcal{N}$ times (layers).\footnote{The $\circ$ denotes composition of functions.}
The final layer $L$ rescales and shifts each component of the output separately:
\begin{align}
    L(\omega_j; \boldsymbol{a}, \boldsymbol{b}) = a_j \, \omega_j + b_j,
\end{align}
with $\boldsymbol{a}$ and $\boldsymbol{b}$ being $N$-dimensional vectors.

The NF is trained in a supervised manner using samples $\{\btheta_i, p(\btheta_i)\}$ from the first MCMC run.
The loss function is given by the Jeffreys' divergence~\eqref{eq:loss_Jeffreys}, which is particularly suitable because it takes its minimum value of zero only when the NF-approximated density $p^\prime(\btheta)$ exactly matches the true posterior $p(\btheta)$, and it does so independently of the (generally unknown) normalization of the posterior (i.e., the evidence).
We will also test using the KL divergence~\eqref{eq:loss_KL} as a loss function in the training of the NF.
For optimization, we use the \textsc{Adam} optimizer~\cite{Kingma:2014vow} with a fixed learning rate, which provides stable convergence during training.

Once trained, the NF provides an efficient and memory-saving way to generate new uncorrelated samples from the approximate posterior, preserving complex correlations even in high-dimensional parameter spaces.
Further details on the NF implementation can be found in Ref.~\cite{Yamauchi:2023xrz}.

\subsection{Unsupervised Learning}
\label{subsubsec:unsupervised}
In addition to the supervised training mode above, which was implemented in the numerical framework~\cite{Yamauchi:2023xrz}, we also consider an unsupervised learning setup to train the normalizing flow.
This option is particularly valuable if only samples generated from a distribution are observed, but not the associated probability densities, for instance, when generating samples from a principal component analysis (PCA) of a multi-dimensional distribution~\cite{Gong:2024lhq}.

Instead of minimizing a loss function, such as Jeffreys' divergence~\eqref{eq:loss_Jeffreys} or the KL divergence~\eqref{eq:loss_KL} based on weighted MCMC samples, the unsupervised method maximizes the following log-likelihood function~\cite{li2022neuralpcaflowbasedrepresentation},
\begin{align}
    \ln(\mathcal{L}) = -\frac{1}{2} \bomega^2 - \frac{1}{2} N \ln(2\pi) + \ln\left[\det\left(\frac{\partial\bomega}{\partial\btheta}\right)\right]
    \label{eq:logL}
\end{align}
with the inverse mapping $\bomega=\mathcal{F}^{-1}(\btheta)$.
In this setting, we treat the empirical sample distribution $p^\prime(\btheta)$ as the actual target and directly train the flow to assign high probability density to these samples.

Rather than using the probability density as the weight for individual samples from an MCMC run, the unsupervised approach trains the NF model to be optimized such that it assigns a high probability to the provided samples.
The training optimizes the likelihood in Eq.~\eqref{eq:logL} by learning the shape of the distribution directly from the samples. Throughout training, the flow samples from the input space and projects it back onto the latent space, where the reference distribution is the multivariate Gaussian. For every sample, the model computes the likelihood of it under the reference distribution.
The effect on the volume of the parameter space of the transformation is then measured by computing the Jacobian determinant of the inverse transformation.
In combination, the two elements -- the volume change and the latent space probability -- enable the model to assign a probability value to the input sample.
The training criterion thus maximizes the aggregate mean likelihood across all samples, or equivalently, minimizes the negative log-likelihood.
This process facilitates the NF model to create a modified distribution that closely approximates the empirical sample distribution without requiring direct familiarity with the data's underlying probability density.

In the next section, we will compare the performance of unsupervised learning with supervised training. 

\subsection{Sequential Bayesian Inference}
\label{subsec:BayesianInference}

Let's consider performing a sequential Bayesian inference study with two sets of experimental measurements, $\{D_1, D_2\}$, that are independent of each other. Starting from Bayes' theorem,
\begin{align}
\mathcal{P}(\btheta|D_1, D_2) &= \frac{\mathcal{P}(D_1, D_2|\btheta)\mathcal{P}(\btheta)}{\mathcal{P}(D_1, D_2)} \nonumber \\
&= \frac{\mathcal{P}(D_2|\btheta) \mathcal{P}(D_1|\btheta)  \mathcal{P}(\btheta)}{\mathcal{P}(D_2) \mathcal{P}(D_1)} \nonumber \\
&=\frac{\mathcal{P}(D_2|\btheta) \mathcal{P}(\btheta|D_1)}{\mathcal{P}(D_2)},
\label{eq:SeqBayes}
\end{align}
where $\mathcal{P}(\btheta|D_1) = \mathcal{P}(D_1|\btheta) \mathcal{P}(\btheta)/\mathcal{P}(D_1)$ is the posterior distribution from applying only experimental dataset $D_1$. The last line in Eq.~\eqref{eq:SeqBayes} shows $\mathcal{P}(\btheta|D_1)$ serves as a prior for the second-stage Bayesian analysis when imposing constraints from dataset $D_2$. Eq.~\eqref{eq:SeqBayes} shows that, mathematically, the sequential Bayesian analysis gives the same posterior distribution as the one-shot analysis by imposing both experimental data together.

In this work, we adopt a seven-dimensional Bayesian inference study from high-energy nuclear physics as an example to explore sequential Bayesian inference. This case study performed a global Bayesian analysis of diffractive $J/\psi$ production in high-energy $\gamma+p$ and $\gamma+\mathrm{Pb}$ collisions with a model framework based on color glass condensate (CGC) theory~\cite{Mantysaari:2025ltq, Mantysaari:2025ujz,Mantysaari:2025cok}.
This study included cross-section measurements from two collision systems: $\gamma+p$ and $\gamma+\mathrm{Pb}$, which were first analyzed separately and then jointly in a combined Bayesian inference~\cite{Mantysaari:2025ltq}. These results obtained at different stages enable us to explore the integration of non-uniform prior distributions in a sequential Bayesian analysis setup. For example, one can use the posterior from the $\gamma+p$ analysis as a prior and then incorporate the constraints from the $\gamma+\mathrm{Pb}$ data, or vice versa. The results from such sequential Bayesian analyses can be directly verified with the combined calibration reported in Ref.~\cite{Mantysaari:2025ltq}. Therefore, our study here provides a way to quantify the consistency and information gain from sequential versus simultaneous analyses.

The seven model parameters and their prior ranges for our case study are listed in Table~\ref {tab:prior_ranges}.
\begin{table}[htb!]
    \caption{Summary of model parameters and their prior ranges~\cite{Mantysaari:2025ltq}.}
    \label{tab:prior_ranges}
    \begin{tabular}{c|c}
    \hline\hline
    Parameter & Prior range \\
    \hline
    $m\;[\mathrm{GeV}]$ &  $[0.02,1.2]$ \\
    $B_G\;[\mathrm{GeV}^{-2}]$ & $[1,10]$ \\
    $B_{q}\;[\mathrm{GeV}^{-2}]$ & $[0.05,3]$ \\
    $\sigma$ & $[0,1.5]$ \\
    $Q_s/(g^2\mu)$ & $[0.05,1.5]$ \\
    $m_{\mathrm{JIMWLK}}\;[\mathrm{GeV}]$ & $[0.02,1.2]$ \\
    $\Lambda_{\mathrm{QCD}}\;[\mathrm{GeV}]$ & $[0.0001,0.28]$ \\
    \hline\hline
    \end{tabular}
\end{table}

In this example, because the theoretical model is computationally expensive, we trained Gaussian Process (GP) emulators~\cite{Rasmussen2006Gaussian} to perform the Bayesian inference study. The emulators in that study are based on the \texttt{surmise} package developed by the BAND collaboration~\cite{surmise2023}, as well as the standard GP implementation provided by the Scikit-learn Python package~\cite{scikit-learn}.

The likelihood function is modeled as a multivariate Gaussian distribution:
\begin{align}
    &\mathcal{P}(\mathbf{y}_{\rm exp}|\btheta) = \frac{1}{\sqrt{2\pi |\mathrm{det}(\Sigma)|}} \nonumber \\
    & ~~~~ \times\exp\left[-\frac{1}{2} (y(\btheta) - y_\mathrm{exp})^\mathsf{T} \Sigma^{-1} (y(\btheta) - y_\mathrm{exp})\right],
\end{align}
where the covariance matrix, $\Sigma = \Sigma_\mathrm{model} + \Sigma_\mathrm{exp}$, accounts for both model and experimental uncertainties. 
In our case, the model uncertainty is estimated from the predictive variance of the GP emulators.
In this work, we assume that different sets of experimental data, namely those from $\gamma+p$ and $\gamma+\mathrm{Pb}$ collisions, are independent of each other, such that the covariance matrix $\Sigma$ is block-diagonal and the likelihood function can be factorized into individual contributions like in Eq.~\eqref{eq:SeqBayes}. 

To sample the posterior distribution, we use the \texttt{pocoMC} package~\cite{Karamanis:2022alw,Karamanis:2022ksp}, which implements advanced Markov Chain Monte Carlo (MCMC) techniques suited for complex, high-dimensional distributions.
The \texttt{pocoMC} sampler attains an efficiency comparable to gradient-based methods such as Hamiltonian Monte Carlo (HMC), while avoiding the need for algorithmic derivatives of the Gaussian-process emulator, which are not provided by the GPs employed in this work.
The emulator and MCMC tools are bundled in the Python package available in Ref.~\cite{hendrik_roch_2025_15879411}, which has been successfully applied in several heavy-ion collision studies~\cite{Roch:2024xhh,Jahan:2024wpj,Gotz:2025wnv,Jahan:2025cbp}. In the next section, we will demonstrate that advanced MCMC sampling techniques are advantageous for our multi-step Bayesian inference by comparing them to the standard \texttt{emcee} MCMC sampler~\cite{Foreman-Mackey:2012any}.

The posterior distributions from Ref.~\cite{Mantysaari:2025ltq} are publicly available in the form of MCMC sample chains of model parameters along with their corresponding log-likelihood values~\cite{mantysaari_2025_15880667}. These outputs enable a supervised training setup, where the parameter vectors serve as inputs and the log-likelihood values as the target probability distribution because the prior is a uniform distribution. We will also apply unsupervised learning by using only the parameter vectors in the samples and compare the performance of the trained NF models from different methods.

\section{Results}
\label{sec:results}
To determine the optimal NF architecture, we perform a hyperparameter scan with $2 \times 10^5$ NF training steps on a dataset of 85k samples. 
We explore combinations of batch sizes $\{500, 1000, 2000, 5000\}$, coupling layers $\mathcal{N} \in \{2, 6, 8, 10, 12\}$, and learning rates $\{1\times 10^{-3}, 5\times 10^{-4}, 1\times 10^{-4}, 5\times 10^{-5}\}$. 
Each trained model generates 85k samples, which are compared to the target distribution using the Kullback-Leibler (KL) divergence defined in Eq.~\eqref{eq:loss_KL} for the 1D marginalized parameter distributions and the 2D pairwise covariance structure.
Finally, the model performance is assessed using the average KL divergence across all dimensions. 
The architecture with the lowest KL divergence is selected for the subsequent Bayesian analysis. 
In the supervised setup, we test two loss functions: Jeffreys’ divergence in Eq.~\eqref{eq:loss_Jeffreys} and the KL divergence in Eq.~\eqref{eq:loss_KL}. 
Unsupervised training, in contrast, always employs the negative log-likelihood objective function.

The NF training setup used for this work is available in Ref.~\cite{hendrik_roch_2025_17122724}.

\subsection{NF Training Results}
\label{subsec:NF_training_results}

\begin{figure*}[htb!]
    \centering
    \includegraphics[width=0.9\linewidth]{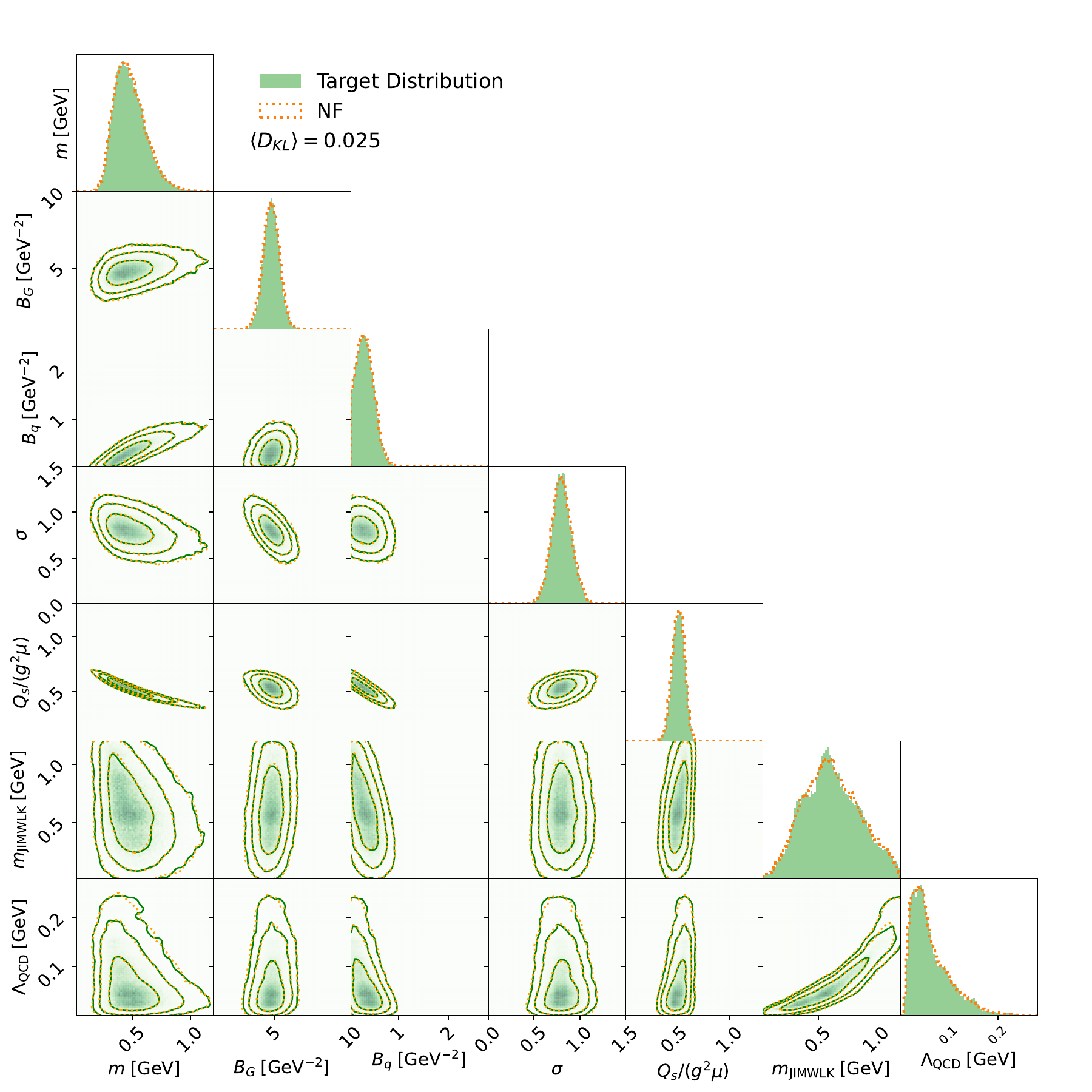}
    \caption{Comparison of the target distribution (green) with samples from the NF model (orange, dotted) for the posterior constrained with the $\gamma+p$ dataset~\cite{Mantysaari:2025ltq}. The NF model utilizes the KL loss function, a batch size of 5000, 6 layers, and a learning rate of $1\times 10^{-3}$. Contour lines indicate the $1\sigma$, $2\sigma$, and $3\sigma$ boundaries.}
    \label{fig:NF_eP}
\end{figure*}
\begin{figure*}[htb!]
    \centering
    \includegraphics[width=0.9\linewidth]{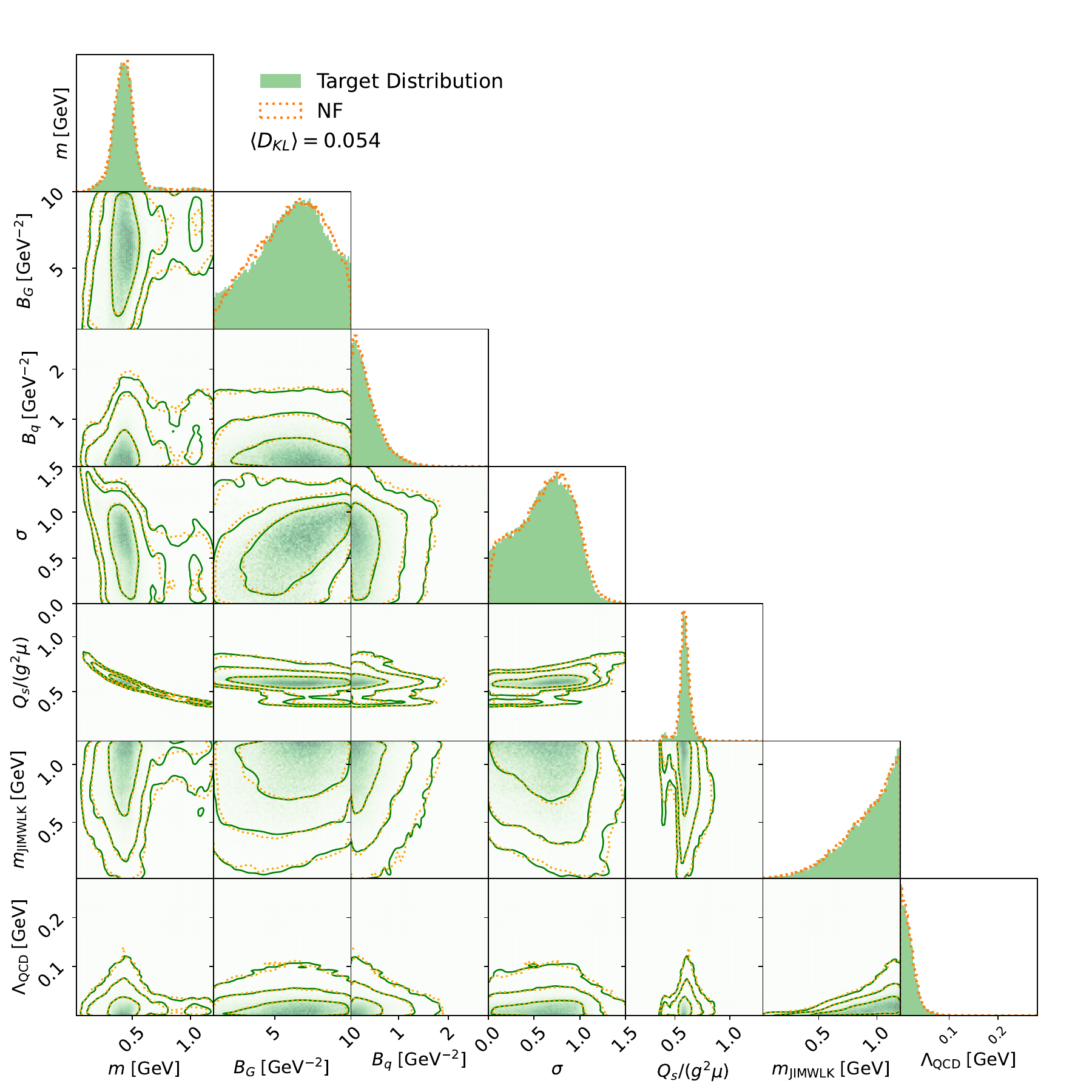}
    \caption{Comparison of the target distribution (green) with samples from the NF model (orange, dotted) for the posterior constrained with the $\gamma+\mathrm{Pb}$ dataset~\cite{Mantysaari:2025ltq}. The NF model utilizes the KL loss function, a training batch size of 1000, 12 layers, and a learning rate of $1 \times 10^{-3}$. Contour lines indicate the $1\sigma$, $2\sigma$, and $3\sigma$ boundaries.}
    \label{fig:NF_ePb}
\end{figure*}

Figure~\ref{fig:NF_eP} compares the NF-generated distributions (orange) with the training posterior samples (green) from imposing the $\gamma+p$ measurements. The agreement is excellent in this case: one-dimensional marginal distributions are mostly unimodal, well within the prior boundaries, and thus easier for the NF to reproduce. The two-dimensional pairwise covariance shows that the NF model captures the nontrivial correlation between parameters in the posterior distribution. 

Figure~\ref{fig:NF_ePb} presents a bigger challenge for the NF model to fit the posterior distribution obtained from the Bayesian inference with the $\gamma+\mathrm{Pb}$ dataset. 
In this case, some parameters ($B_G$, $\sigma$) have broad distributions, while others ($B_q$, $m_\mathrm{JIMWLK}$, and $\Lambda_{\rm QCD}$) show peaks near the prior boundaries with long tails. 
Such features challenge the NF to find a mapping to transform them into a Gaussian distribution. 
As shown in Fig.~\ref{fig:NF_ePb}, deviations appear for $B_G$ near prior edges. However, the NF captures boundary peaks in $B_q$, $m_{\rm JIMWLK}$, and $\Lambda_{\rm QCD}$ and reproduces the heavy-tailed $m$ distribution reasonably well.

We compute the averaged KL divergence over all dimensions to quantify the quality of the fit globally, and list the best-performing NF configurations for the $\gamma+p$ and $\gamma+\mathrm{Pb}$ datasets in Table~\ref{tab:KL_divergences}. 
\begin{table}[htb!]
    \caption{Best NF model configurations and corresponding average KL divergence $\langle D_{\rm KL} \rangle$ for the $\gamma+p$ and $\gamma+\mathrm{Pb}$ datasets.}
    \label{tab:KL_divergences}
    \begin{tabular}{c|c|c|c|c|c}
    \hline\hline
    Dataset & Batch Size & Layers & Learning Rate & Loss & $\langle D_{\rm KL}\rangle$ \\
    \hline
    $\gamma+p$ & 500 & 4 & $1\times 10^{-4}$ & Jeffreys' & 0.047 \\
    $\gamma+p$ & 5000 & 6 & $1\times 10^{-3}$ & KL & 0.025 \\
    $\gamma+p$ & 1000 & 10 & $1\times 10^{-3}$ & log-$\mathcal{L}$ & 0.024 \\
    \hline
    $\gamma+\mathrm{Pb}$ & 500 & 12 & $1\times 10^{-3}$ & Jeffreys' & 0.061 \\
    $\gamma+\mathrm{Pb}$ & 1000 & 12 & $1\times 10^{-3}$ & KL & 0.052 \\
    $\gamma+\mathrm{Pb}$ & 2000 & 10 & $1\times 10^{-3}$ & log-$\mathcal{L}$ & 0.052 \\
    \hline\hline
    \end{tabular}
\end{table}
The average KL divergence $\langle D_\mathrm{KL}\rangle$ between the trained NF models and the target posterior distributions is quite small for all three setups listed in Tab.~\ref{tab:KL_divergences}.
The number of training layers in the optimized NF models for $\gamma+\mathrm{Pb}$ data is noticeably larger than those needed to fit the $\gamma+p$ data, reflecting that it is more challenging to capture all the features in the posterior distribution from the $\gamma+\mathrm{Pb}$ collisions. 
For both experimental datasets, the KL divergence loss function consistently outperforms Jeffreys’ divergence, potentially because we use the average KL divergence as a quality measure.
Interestingly, unsupervised training with the log-likelihood achieves accuracy comparable to that of models using the KL loss. However, without conducting a thorough test of unsupervised training across different distributions, we do not know whether it performs as well as the supervised approach in general.

In our current study, we expect to obtain very similar results from both approaches, as the two training methods yield comparable performance. For simplicity, we adopt the supervised NF models trained with the KL loss for both datasets for the remainder of this study.

\subsection{Multi-Stage Bayesian Inference}
\label{subsec:multi_stage_Bayesian_inference}
\begin{figure*}[htb!]
    \centering
    \includegraphics[width=0.9\linewidth]{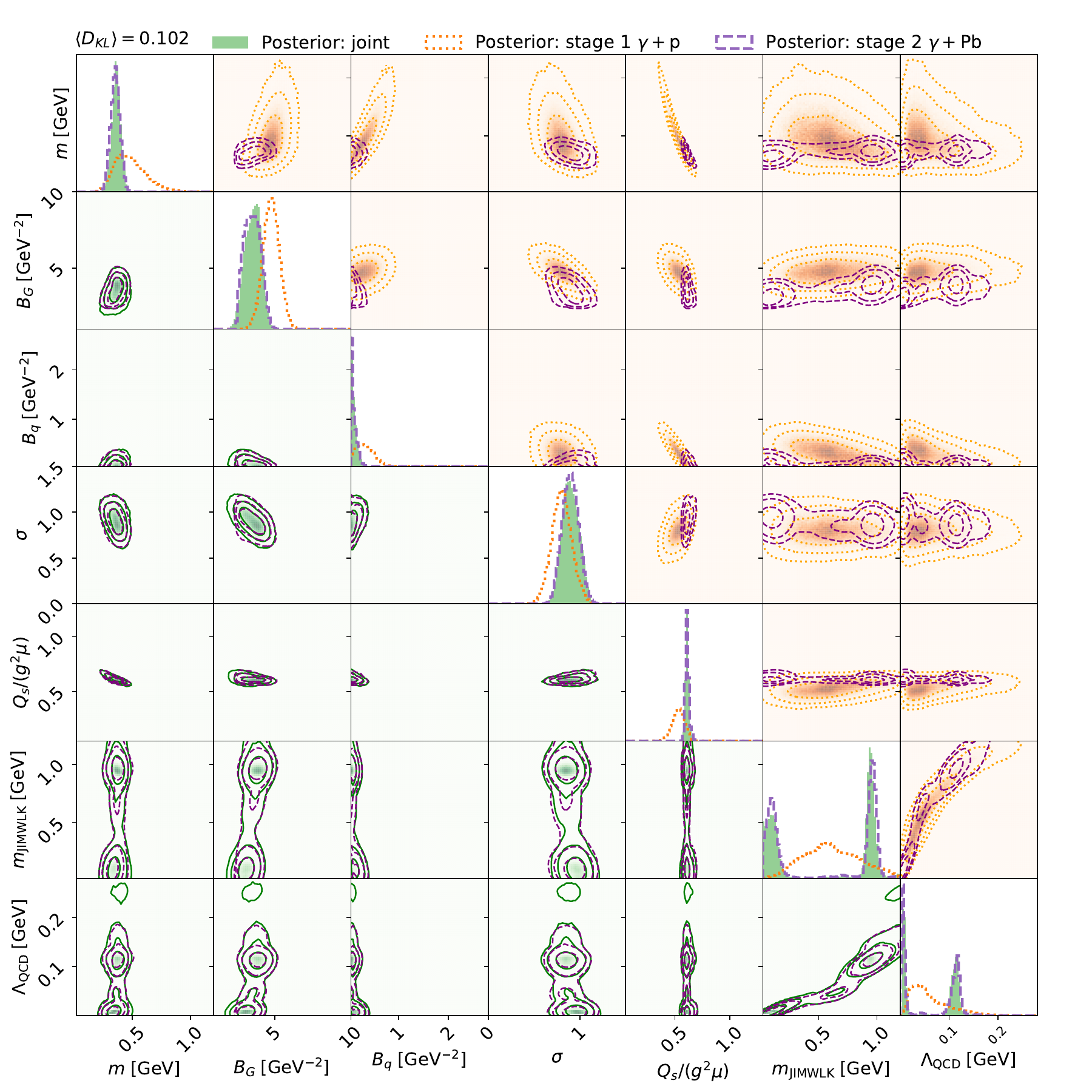}
    \caption{Multi-stage Bayesian inference starting from the posterior constrained with the $\gamma+p$ data, then inference with the $\gamma+\mathrm{Pb}$ data. Full lines (green) indicate the joint inference, dotted lines (orange) the first-stage posterior, and dashed lines (purple) the second-stage posterior. Contours show $1\sigma$, $2\sigma$, and $3\sigma$ boundaries.}
    \label{fig:MCMC_eP_prior}
\end{figure*}

\begin{figure*}[htb!]
    \centering
    \includegraphics[width=0.9\linewidth]{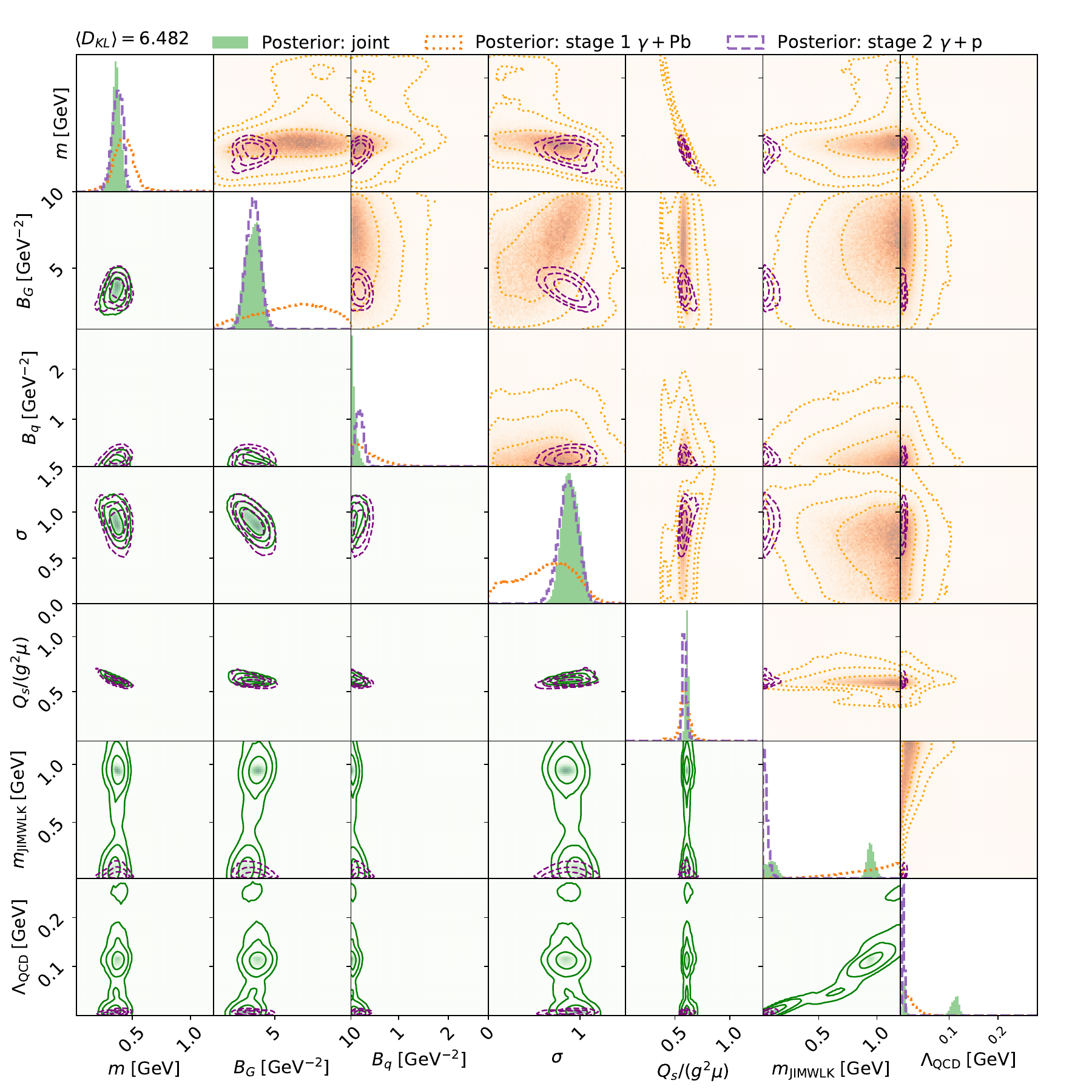}
    \caption{Multi-stage Bayesian inference starting from the posterior constrained by the $\gamma+\mathrm{Pb}$ data, then inference with the $\gamma+p$ data. Full lines (green) indicate the joint inference, dotted lines (orange) the first-stage posterior, and dashed lines (purple) the second-stage posterior. Contours show $1\sigma$, $2\sigma$, and $3\sigma$ boundaries.}
    \label{fig:MCMC_ePb_prior}
\end{figure*}

\begin{figure*}[htb!]
    \centering
    \includegraphics[width=0.9\linewidth]{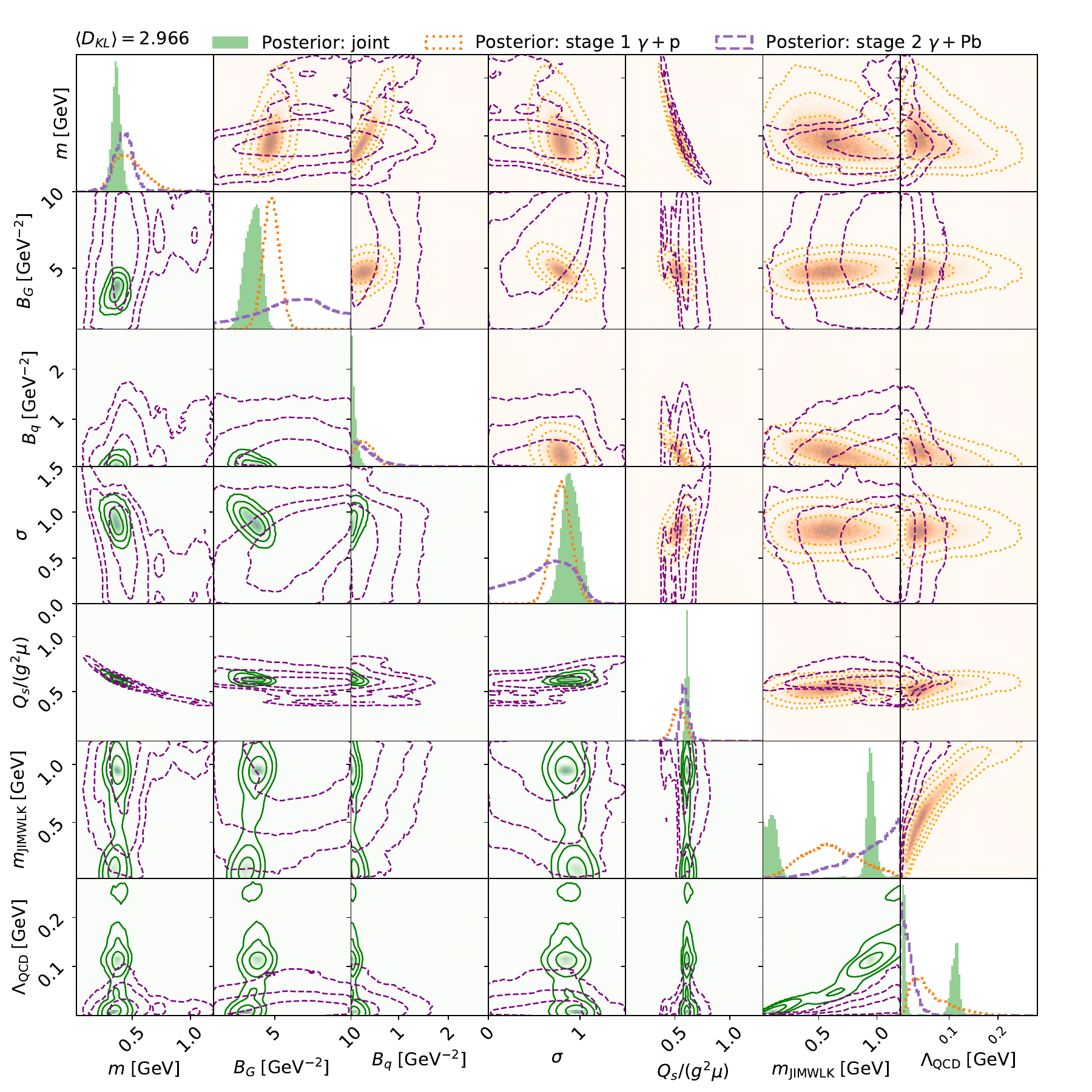}
    \caption{Same inference as Fig.~\ref{fig:MCMC_eP_prior}, but replacing the second-stage \texttt{pocoMC} sampler with \texttt{emcee}. Contours show $1\sigma$, $2\sigma$, and $3\sigma$ boundaries.}
    \label{fig:MCMC_eP_prior_emcee}
\end{figure*}

We now turn to multi-stage Bayesian inference setups. 
The reference posterior from the joint analysis of $\gamma+p$ and $\gamma+\mathrm{Pb}$ data, obtained in Ref.~\cite{Mantysaari:2025ltq}, is shown as solid green contours in the following corner plots (Figs.~\ref{fig:MCMC_eP_prior}-\ref{fig:MCMC_eP_prior_emcee}). 
In the sequential Bayesian analyses, starting from either dataset, the first-stage posterior (orange, dotted contours) is learned by the NF model and then used as a prior for the second-stage inference, whose posterior distribution is plotted as purple dashed contours. 
The upper-right corner plots show how the posterior distribution changes from stage 1 to stage 2 in the sequential Bayesian analysis. The lower-left corner plots compare the final posterior distribution obtained from the sequential Bayesian analysis with the reference distribution from a one-shot joint Bayesian inference with both sets of data.
We compare the contours of the distributions at $1\sigma$, $2\sigma$, and $3\sigma$ levels. 

Figure~\ref{fig:MCMC_eP_prior} shows the results when starting from $\gamma+p$ data in the sequential Bayesian analysis. 
In this case, the first-stage posterior is broader than the second-stage posterior, reflecting the additional constraints introduced by the $\gamma+\mathrm{Pb}$ data.
The multi-stage approach reproduces the joint posterior well, with an average KL divergence $\langle D_\mathrm{KL} \rangle \approx 0.1$. We note that the multi-modal structures in $m_{\rm JIMWLK}$ and $\Lambda_{\rm QCD}$ are well reproduced in the sequential Bayesian analysis. 
Here, the wide coverage from the first stage aids the exploration of parameter space when potential multi-modal structures are present in the final result.

Figure~\ref{fig:MCMC_ePb_prior} explores the inverse order in the sequential Bayesian analysis, starting from $\gamma+\mathrm{Pb}$ data. 
In this case, the first-stage posterior distribution is broad enough in the first six dimensions compared to the final posterior. However, the $\gamma+\mathrm{Pb}$ data favors small values of $\Lambda_{\rm QCD}$ parameters, leaving a very small probability for $\Lambda_{\rm QCD} \approx 0.1$ GeV. Consequently, the second-stage posterior cannot reproduce the multi-modal structure seen in the full calibration. The MCMC sampler has difficulty exploring the model parameter phase space around $\Lambda_{\rm QCD} \approx 0.1$ GeV in the second stage. This result highlights one limitation of the multi-stage Bayesian approach: when the first-stage posterior misses relevant modes, they cannot be recovered in later stages. In this case, the final posterior from the sequential Bayesian analysis yields a much larger KL divergence $\langle D_{\rm KL} \rangle = 6.482$ compared to the results in Fig.~\ref{fig:MCMC_eP_prior}.

These results underscore the need for caution in multi-stage Bayesian inference, especially when multi-modal structures are present in the posterior distribution, or there is tension for the theoretical model to reproduce different experimental datasets with a single set of parameters.
In practical applications, however, one often begins with broad, low-statistics observables before incorporating more constraining data. In such cases, the iterative approach should remain effective.

Finally, Fig.~\ref{fig:MCMC_eP_prior_emcee} shows the results of using the standard \texttt{emcee} MCMC sampler for the second stage of Bayesian inference, starting with the NF model fitted to the posterior distribution constrained by $\gamma+p$ data. In this case, the posterior distribution from the second stage of Bayesian inference with the $\gamma + \mathrm{Pb}$ data completely failed to reproduce the reference posterior distribution obtained from one-shot joint analysis.
The different performance in Figs.~\ref{fig:MCMC_eP_prior} and \ref{fig:MCMC_eP_prior_emcee} underscores the importance of robust MCMC sampling in reproducing joint posteriors, particularly for multi-modal distributions. Our study would serve as a useful benchmark for future performance comparisons of other MCMC samplers, such as those based on the HMC approach.

In the Appendix, we present a simplified example where the $\gamma+\mathrm{Pb}$ dataset is split into integrated and differential cross sections. 
In this case, the posterior distribution exhibits no bimodal structures, allowing the multi-stage inference to reproduce the joint posterior much more accurately and independently of the calibration order.

\section{Conclusion}
\label{sec:conclusion}
In this work, we have explored the use of normalizing flows as a flexible tool for constructing informed priors in Bayesian inference for high-dimensional parameter spaces. By training the NF models on posteriors from previous analyses, we demonstrated that they are capable of accurately reproducing complex features of these distributions, including non-Gaussian shapes, correlations, and boundary effects.
We compared supervised and unsupervised training strategies and found that trained NF models with the KL loss function provide the most accurate reproductions. Unsupervised training based on maximum likelihood offers a promising alternative when posterior weights are unavailable.  

Applying these trained NF models as priors in sequential Bayesian workflows, we show that they preserve consistency with reference posteriors obtained from simultaneous fits, while providing a practical framework for reusing information across different datasets. 
This approach, therefore, provides a systematic method for incorporating prior knowledge into future analyses without relying on overly simplified assumptions, such as uniform or Gaussian priors.  

Looking ahead, the method can be extended to more complex applications in high-energy nuclear physics and beyond, where iterative Bayesian analyses are necessary and computational cost is a limiting factor. 
In particular, integrating NF-based priors with advanced MCMC samplers opens the door to significant efficiency gains in large-scale inference studies. 
Further work will focus on deploying this framework to real sequential Bayesian analysis in high-energy nuclear physics.  

Overall, our results highlight the potential of normalizing flow models to serve as powerful and reusable building blocks for Bayesian inference, enabling more efficient and informed exploration of theoretical models in nuclear and particle physics.

\begin{acknowledgments}
We thank Kyle Godbey, Sunil Jaiswal, and Haydar Mehryar for fruitful discussions.
This work is supported in part by the U.S. Department of Energy, Office of Science, Office of Nuclear Physics, under DOE Award No.~DE-SC0021969 and DE-SC0024232. H.~R. and C.~S. were supported in part by the National Science Foundation (NSF) within the framework of the JETSCAPE collaboration (OAC-2004571).
C.S. acknowledges a DOE Office of Science Early Career Award. 
Numerical simulations presented in this work were performed at the Wayne State Grid, and we gratefully acknowledge their support.
\end{acknowledgments}

\appendix*
\section{A simplified example for multi-stage Bayesian analysis}
\label{app:simple_example}

In this Appendix, we present a simplified setup using only the $\gamma+\mathrm{Pb}$ dataset, separated into integrated (int.) and $t$-differential (diff.) cross-section parts~\cite{Mantysaari:2025ltq}. This decomposition avoids the bimodal structure present in the full analysis, providing a simpler test case for the normalizing flow (NF) training and multi-stage Bayesian inference.

Table~\ref{tab:KL_divergences_ePb} lists the best-performing NF configurations for the two datasets across different loss functions.  
Consistent with the main text, we find that the KL loss function outperforms Jeffreys', while the log-likelihood loss achieves results comparable to the KL case.
\begin{table}[htb!]
    \caption{Best NF model configurations and corresponding average KL divergence $\langle D_{\rm KL} \rangle$ for the integrated and differential $\gamma+\mathrm{Pb}$ datasets.}
    \label{tab:KL_divergences_ePb}
    \begin{tabular}{c|c|c|c|c|c}
    \hline\hline
    Dataset & Batch Size & Layers & Learning Rate & Loss & $\langle D_{\rm KL}\rangle$ \\
    \hline
    int. & 2000 & 12 & $1\times 10^{-3}$ & Jeffreys' & 0.054 \\
    int. & 5000 & 8 & $1\times 10^{-3}$ & KL & 0.049 \\
    int. & 5000 & 12 & $1\times 10^{-3}$ & log-$\mathcal{L}$ & 0.048 \\
    \hline
    diff. & 2000 & 12 & $1\times 10^{-3}$ & Jeffreys' & 0.063 \\
    diff. & 2000 & 10 & $1\times 10^{-3}$ & KL & 0.061 \\
    diff. & 5000 & 10 & $1\times 10^{-3}$ & log-$\mathcal{L}$ & 0.059 \\
    \hline\hline
    \end{tabular}
\end{table}

\begin{figure*}[htb!]
    \centering
    \includegraphics[width=0.9\linewidth]{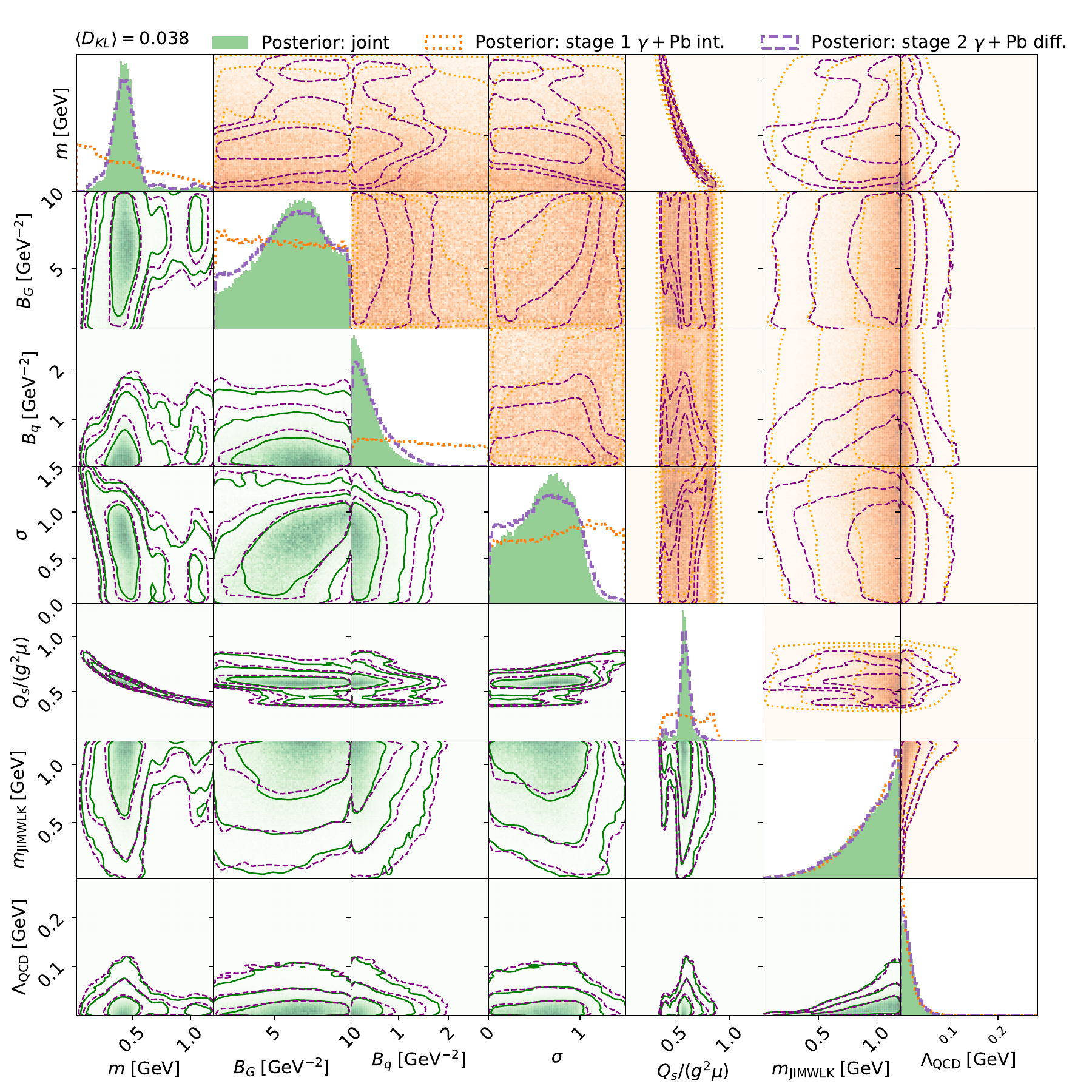}
    \caption{Two-stage Bayesian inference starting from the posterior constrained with $\gamma+\mathrm{Pb}$ integrated dataset and then inference with the $t$-differential cross section dataset in the second stage.  
    Full lines (green) indicate the joint-inference posterior, dotted lines (orange) show the first-stage posterior, and dashed lines (purple) the second-stage posterior.  
    Contours mark the $1\sigma$, $2\sigma$, and $3\sigma$ levels.}
    \label{fig:MCMC_eP_int_prior}
\end{figure*}

Using the NF trained with the KL loss function, we then perform the two-step Bayesian inference and compare the resulting posteriors with the reference obtained from the one-shot joint inference.  
Figure~\ref{fig:MCMC_eP_int_prior} shows the sequential Bayesian setup starting from the integrated cross-section dataset.  
The first-stage posterior (dotted orange) is relatively broad for the first four parameters, while $Q_s/(g^2\mu)$ exhibits a plateau near the prior center.  
The $m_{\rm JIMWLK}$ and $\Lambda_{\rm QCD}$ parameters peak near the edges of the prior but retain broad tails.  
Including the differential dataset in the second stage leads to significant additional constraints, closely reproducing the joint-inference posterior with an average divergence of $\langle D_{\rm KL}\rangle = 0.038$.

\begin{figure*}[htb!]
    \centering
    \includegraphics[width=0.9\linewidth]{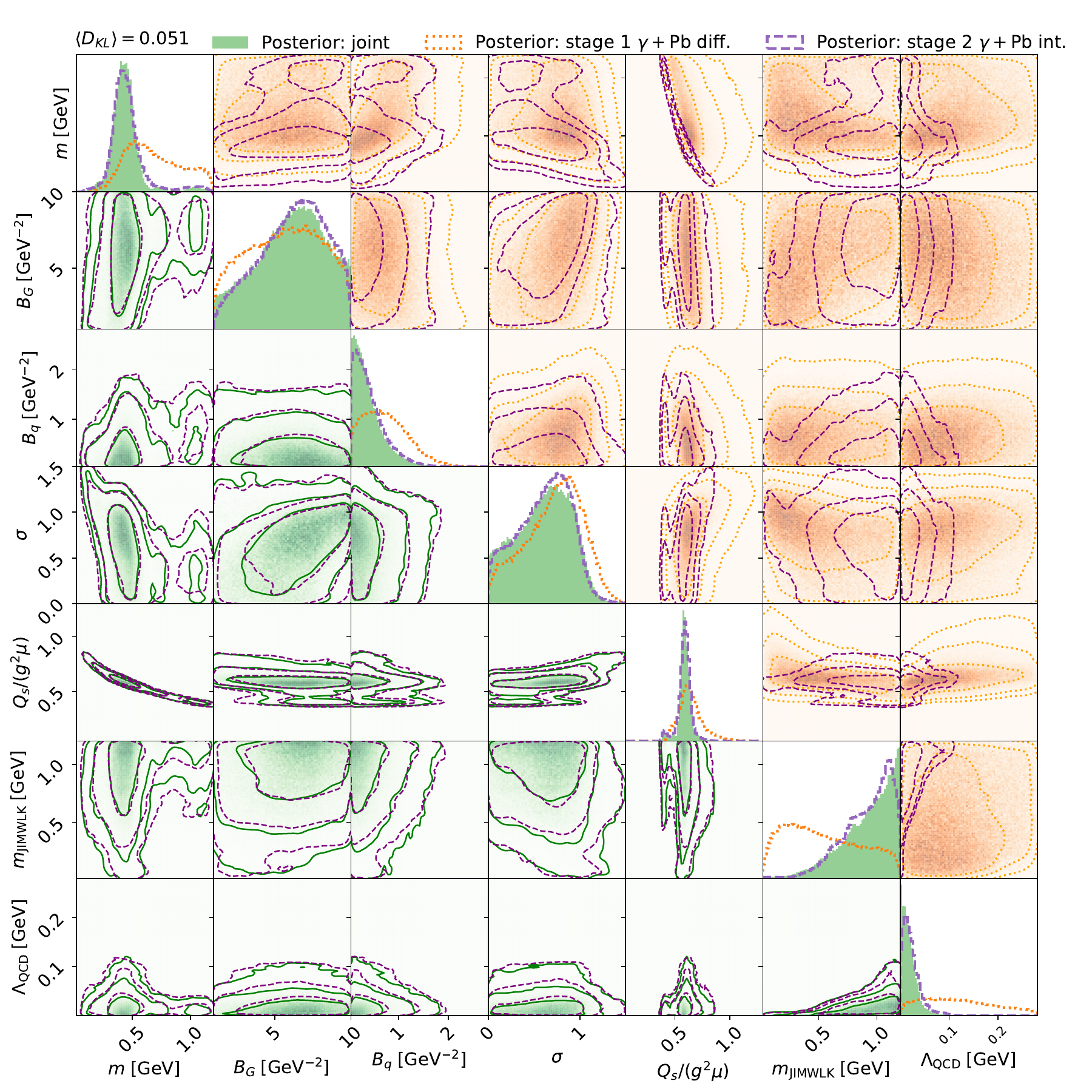}
    \caption{Two-stage Bayesian inference starting from the posterior constrained with the $\gamma+\mathrm{Pb}$ $t$-differential cross sections and inference with the integrated cross section dataset in the second stage.  
    Full lines (green) indicate the joint-inference posterior, dotted lines (orange) show the first-stage posterior, and dashed lines (purple) the second-stage posterior.  
    Contours mark the $1\sigma$, $2\sigma$, and $3\sigma$ levels.}
    \label{fig:MCMC_ePb_diff_prior}
\end{figure*}

Figure~\ref{fig:MCMC_ePb_diff_prior} presents the sequential Bayesian setup with the reverse order, starting with the constraints from the $t$-differential cross-section dataset at the first stage.
Again, the broad first-stage posterior can be refined in the second stage, yielding results close to the joint calibration with $\langle D_{\rm KL}\rangle = 0.051$.

This simplified example demonstrates that both inference orders are viable and can achieve posteriors comparable to those obtained through joint calibration when no multi-modal structures are present.  
For completeness, we also verified that the \texttt{emcee} MCMC sampler still fails to reproduce the joint results when applied in the second stage.

\bibliography{bib, non-inspire}

\end{document}